\documentclass{article}
\usepackage{fullpage}
\usepackage{listings}
\usepackage{natbib}
\usepackage{graphicx}
\usepackage{hyperref}
\usepackage{microtype}
\usepackage{graphicx}
\usepackage{soul, color}
\usepackage{xcolor}
\usepackage{siunitx}
\usepackage{authblk}
\usepackage{subfigure}
\usepackage{amsfonts}
\usepackage{amsmath}
\usepackage{mathtools}
\usepackage[noend]{algpseudocode}
\usepackage{graphics}
\usepackage[top=1in, bottom=1in, left=1in, right=1in]{geometry}    

\usepackage{authblk}
\title{Active Learning to Classify Macromolecular Structures \emph{in situ} for Less Supervision in Cryo-Electron Tomography }

\author[1]{Xuefeng Du}
\author[2]{Haohan Wang}
\author[4]{Zhenxi Zhu}
\author[3]{Xiangrui Zeng}
\author[5]{Yi-Wei Chang}
\author[6]{Jing Zhang}
\author[3,*]{Min Xu}
\affil[1]{Department of Computer Science, University of Wisconsin-Madison, Madison, 53706, USA}
\affil[2]{Language Technologies Institute, Carnegie Mellon University, Pittsburgh, 15213, USA}
\affil[3]{Computational Biology Department, Carnegie Mellon University, Pittsburgh, 15213, USA}
\affil[4]{Department of Computer Science and Technology, Nanjing University, 210023, China}
\affil[5]{Department of Biochemistry and Biophysics, University of Pennsylvania, Philadelphia, 19104, USA}
\affil[6]{Department of Computer Science, University of California - Irvine, Irvine, 92697, USA}
\affil[*]{Corresponding author}
\date{}

\begin{document}

\maketitle

\abstract{\textbf{Motivation:} Cryo-Electron Tomography (cryo-ET) is a 3D bioimaging tool that visualizes the structural and spatial organization of macromolecules at a near-native state in single cells, which has broad applications in life science. However, the systematic structural recognition and recovery of macromolecules captured by cryo-ET are difficult due to high structural complexity and imaging limits. Deep learning based subtomogram classification have played critical roles for such tasks. As supervised approaches, however, their performance relies on sufficient and laborious annotation on a large training dataset.\\
\textbf{Results:} To alleviate this major labeling burden, we proposed a Hybrid Active Learning (HAL) framework for querying  subtomograms for labelling from a large unlabeled subtomogram pool. Firstly, HAL adopts uncertainty sampling to select the subtomograms that have the most uncertain predictions. This strategy enforces the model to be aware of the inductive bias during classification and subtomogram selection, which satisfies the \emph{discriminativeness} principle in AL literature. Moreover, to mitigate the sampling bias caused by such strategy, a discriminator is introduced to judge if a certain subtomogram is labeled or unlabeled and subsequently the model queries the subtomogram that have higher probabilities to be unlabeled. Such query strategy encourages to match the data distribution between the labeled and unlabeled subtomogram samples, which essentially encodes the \emph{representativeness} criterion into the subtomogram selection process. Additionally, HAL introduces a subset sampling strategy to improve the diversity of the query set, so that the information overlap is decreased between the queried batches and the algorithmic efficiency is improved.
Our experiments on subtomogram classification tasks using both simulated and real data demonstrate that we can achieve comparable testing performance (on average only 3\% accuracy drop) by using less than $30\%$ of the labeled subtomograms, which shows a very promising result for subtomogram classification task with limited labeling resources. \\
\textbf{Availability:} https://github.com/xulabs/aitom \\
\textbf{Contact:} {mxu1@cs.cmu.edu}\\
}

\maketitle
\section{Introduction}
Cellular processes are generally governed by macromolecules. To accurately understand these processes, Cryo-Electron Tomography (cryo-ET) has been developed recently to enable a systematic 3D visualization of subcellular structures in single cells at sub-molecular resolution and in native state. However, due to the structural content complexity of the captured tomograms and imaging limitations, it is difficult to classify macromolecules in subtomograms (A \emph{subtomogram} is a subvolume of a tomogram that is likely to contain a single macromolecule) for structural recovery via manual inspections. Given that subtomogram classification is essentially a 3D image classification problem, supervised deep learning has recently become a major approach thanks to its ability to extract complex image composition rules from big image data. However, even though different approaches have been developed on either 2D or 3D cryo-ET data \citep{DBLP:journals/mva/CheLZEGX18,DBLP:journals/bioinformatics/XuCMLYZX17,DuZZSX19,LiuDXXZZX19}, few of them emphasize the labeling burden, which is very time-consuming and requires structural biology expertise. This situation impedes the off-the-shelf deployment of these algorithms. For instance, even for just 1,000 real subtomograms were used in \citep{DBLP:conf/bmvc/LiuZWGX18}, it already introduced a time-consuming labeling work for domain experts. 

\begin{figure}[t]
	\centering  
	\includegraphics[width=0.8\linewidth] {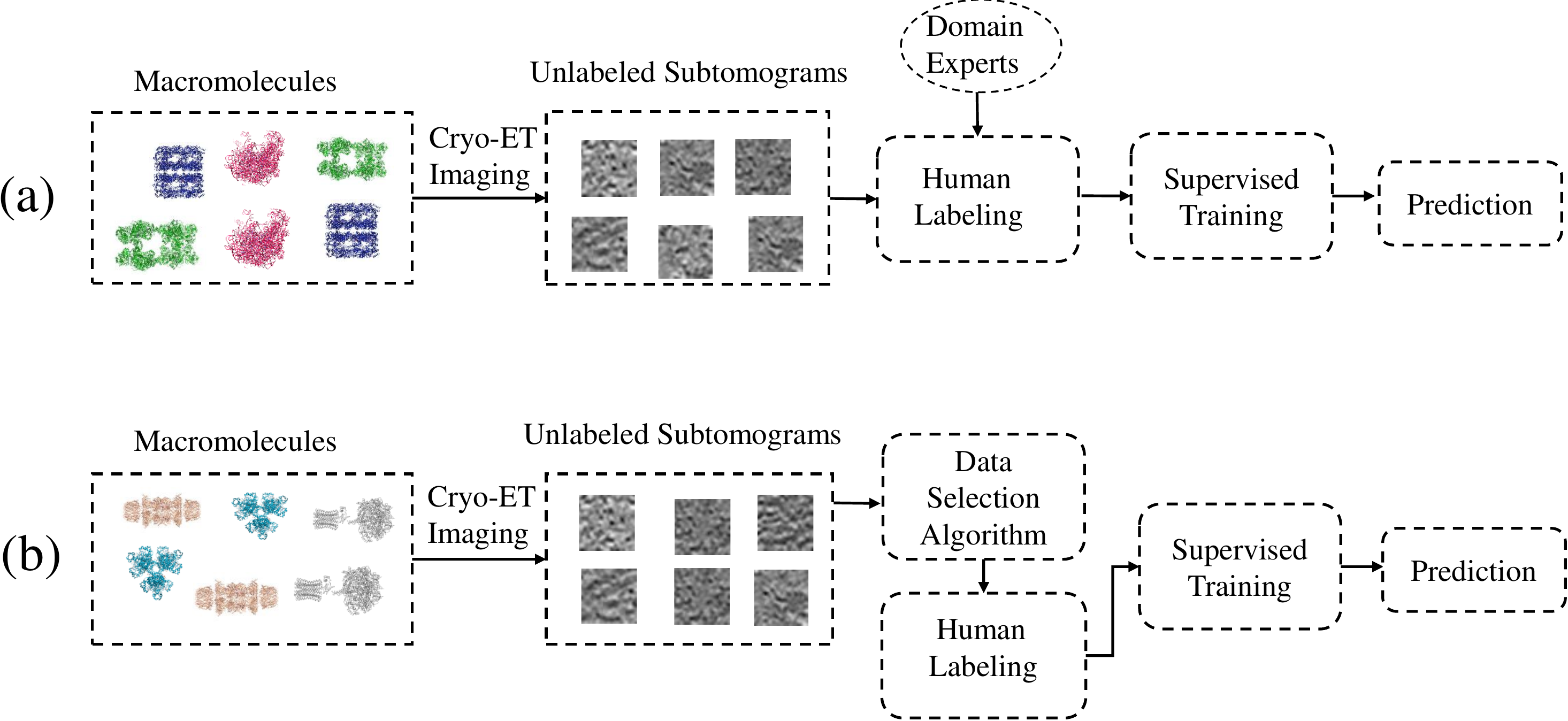}  
	\caption{Exemplary illustration of the active learning approach in cryo-ET classification. (a) shows the existing passive recognition pipeline where sufficient labeling is required where (b) demonstrates the recognition scenario guided by active learning. }
	\label{fig:overview}
\end{figure}

Under such circumstances, we resort \emph{active learning}, which selects a subset of subtomogram samples, if labeled and used for training, will best improve the model's performance under the same labeling budget \citep{Sener2018Active,abs-1907-06347} (Fig.\ref{fig:overview}). Two main principles for such unlabeled sample selection are proposed \citep{DBLP:journals/tcs/Dasgupta11} and they both have limitations: \emph{discriminativeness} and \emph{representativeness}. The discriminativeness principle aims to find the most discriminative samples for the current classifier, which will shrink the space of candidate classifiers as rapidly as possible \citep{DBLP:conf/kdd/WangY13}. The popular proposed criteria are uncertainty rule \citep{Yang2018A,WangCSYS18}, expected error reduction \citep{DBLP:journals/corr/HuangCRLSC16} and query by committee \citep{SeungOS92,Gilad_BachrachNT05}. In this case, the samples are selected based on specific criterion instead of being i.i.d. sampled. Such sampling bias prevents active learning from finding a classifier with good generalization performance and query efficiency \citep{DBLP:conf/kdd/WangY13}, which becomes even severe for high-dimensional and complex 3D medical images. The representativeness principle aims to address this problem by querying the samples which can represent the overall patterns or statistics of the unlabeled data, such as by clustering \citep{NguyenS04} and generative models \citep{zhu2017generative,tran2019bayesian,lee2019bald,SinhaED19,abs-2002-04709}. Such methods perform better when fewer initial labeled data is provided. However, they will become inefficient with the increase of queried classes, as they solely rely on data distributions and do not fully use the label information \citep{DBLP:conf/kdd/WangY13}.

\begin{figure}[tb]
	\centering  
	\includegraphics[width=0.8\linewidth] {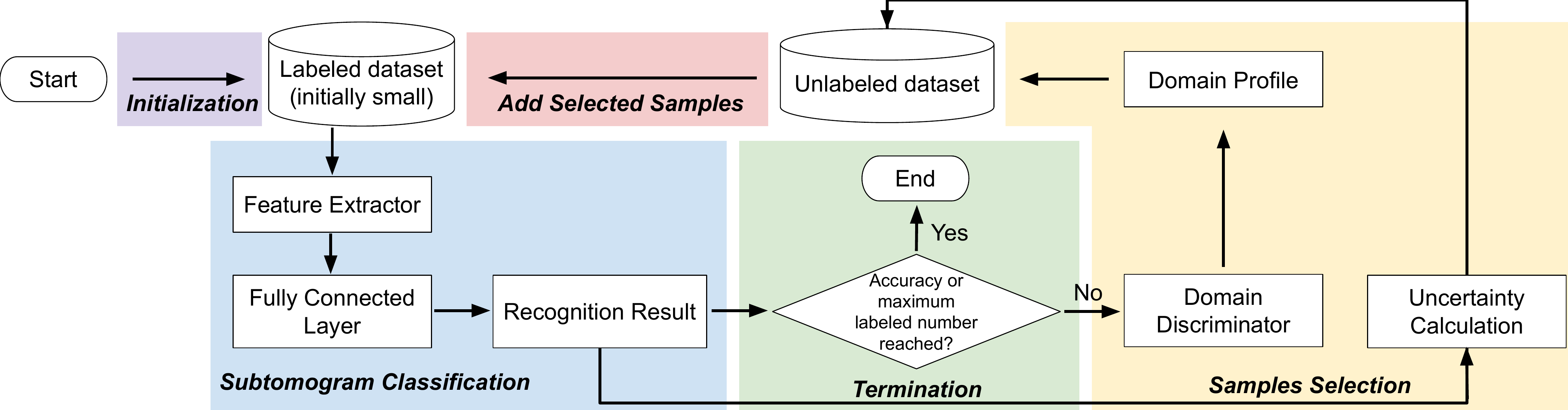}  
	\caption{The full active learning scheme for subtomogram classification in our HAL framework.} 
	\label{fig:flowchart}
\end{figure}

Since using either type of principle alone is not enough to guarantee the optimal result, in this paper, we approach this task by integrating the discriminativeness and representativeness in one optimization formulation, namely the Hybrid Active Learning (HAL) framework. To satisfy the principle of data representativeness, we start with a small labeled set and a large unlabeled set and train a supervised Convolutional Neural Network (CNN) on the labeled set. We then extract the feature representations of both the labeled and unlabeled set. Inspired by the distribution alignment techniques \citep{DBLP:conf/icml/GaninL15}, in each iteration, we train a discriminator on these representations and predict how likely each subtomogram sample is labeled or unlabeled. Then, we select and label those subtomogram samples in the unlabeled dataset which is predicted to have higher probabilities of coming from unlabeled dataset. This alternative optimization scheme effectively improves the representativeness of the labeled training set. Moreover, since the subtomograms captured by cryo-ET are highly heterogeneous, a large selected batch is likely to contain redundant subtomogram samples, which leads to a significant information overlap and thus an inefficient querying process. Therefore, we apply a sub-sampling strategy to enlarge the query batch without losing diversity. For the discriminativeness principle, We additionally introduce the label information by using the entropy of predictions as selection criterion. Such heuristic is a strong active learning baseline, namely uncertainty sampling \citep{Yang2018A}. In each sampling iteration, we use both principles to score the current unlabeled subtomogram samples and then ensemble the two scores for final ranking and selection. We then add all queried subtomogram samples into the labeled dataset and repeat until the labeling budget is reached. The overall learning and querying steps are summarized in Figure \ref{fig:flowchart}. Note that the hybrid querying heuristics are also proposed in literature \citep{YinQCLWZD17,AshZK0A20} and we defer the discussion in Section~\ref{sec:active_learning}.
The contributions of this paper is summarized as follows: 1) We propose a 3D HAL framework to query unlabeled subtomogram samples and expand the training dataset, such that deep models can be trained with significantly lower labeling cost while incurring minimal prediction accuracy drop. We provide a theoretical analysis of the expected classification risk of our framework (Equation \ref{eq:proof}). 2) HAL is the first active learning work to address the issue of labeling cost in cryo-ET analysis tasks, which integrates two principled query heuristics in one optimization framework to make the queried subtomograms both representative and discriminative. 3) In HAL, we adopt several effective strategies to improve the performance, such as proposing a convolutional discriminator to learn the comparative metric of representations from shallower layers, introducing sub-sampling to improve the diversity of every query batch. 4) The empirical results for subtomogram classification using both simulated and real data demonstrate that we are able to achieve comparable testing performance (on average only $<$ 3\% prediction accuracy drop) while significantly reducing the labeling burden by over 70\%. 

\section{Related Works}
\subsection{Deep Learning based cryo-ET Analysis}
Since 2017, supervised deep learning started getting popular for cryo-ET analysis. \cite{chen2017convolutional} proposed a 2D CNN segmentation model for segmenting ultrastructures on 2D slices of a 3D tomogram. Supervised deep learning based methods have also been proposed for 3D subtomogram classification task thanks to its high-throughput processing capability. Specifically, \citep{DBLP:journals/bioinformatics/XuCMLYZX17} was the first to propose a deep learning approach to separate different structures in subtomograms. \cite{DBLP:conf/bmvc/LiuZWGX18} introduced multi-task learning to learn a feature space for simultaneous classification, segmentation, and structural recovery of macromolecules. Other techniques, such as domain adaptation \citep{abs-2007-15422}, open-set recognition \citep{DuZZSX19} and semi-supervised learning \citep{LiuDXXZZX19} have also been applied for reliable and deployable analysis result. Other works are~\cite{LiuZLLFXX18} and . However, few of them take the labeling burden into account, which provided a time-consuming labeling work for domain experts and impeded their off-the-shelf usage. 

\subsection{Active Learning}
\label{sec:active_learning}
Active learning has been popular prior to deep learning. It was usually used on small models. The uncertainty-based methods usually measure uncertainty by posterior probability of the predicted classes \citep{LewisC94,Lewis95a} or the margin between the first predicted class and the second confidently predicted class \citep{JoshiPP09,RothS06}. Other methods measure uncertainty via entropy \citep{SettlesC08,LuoSU13,JoshiPP09} or the distance to the decision boundary for SVM \citep{LiG14,TongK01,VijayanarasimhanG11a}. Another direction that is based on the discriminative principle, such as query by committee, usually develops multiple models as a committee and uses the disagreement between them as the criteria for uncertainty estimation \citep{McCallumN98,SeungOS92}. The representativeness principle selects samples that can cover the distribution of the entire dataset by clustering \citep{NguyenS04}, discrete optimization \citep{YangMNCH15,Guo10,ElhamifarSYS13}. Other popular methods either focus on the neighboring information between samples \citep{0003R15,AodhaCKB15,XuKJ10} or the expected model change \citep{SettlesCR07,RoyM01,FreytagRD14} for sample selection.

When deep learning comes in, several query heuristics are proposed on larger models and datasets, such as uncertainty based methods \citep{LinWMZZ18,WangZLZL17,Gal2017Deep,BeluchGNK18}, core-set sampling \citep{Sener2018Active}, generative models \citep{SinhaED19}. However, some of them, especially core-set sampling, are not scalable on very large datasets compared to our HAL because a large distance matrix from unlabeled samples is needed which makes the query procedure highly expensive. Some hybrid query approaches are also proposed. For instance, \cite{YinQCLWZD17} selected the data by uncertainty sampling and random sampling while we use a discriminator to query representative samples in order to reduce sampling bias in uncertainty sampling. \cite{AshZK0A20} selected samples whose gradients span a diverse directions, which did not explicitly consider different query heuristics. \cite{ShuiZGW20} used the Wasserstein metric for measuring the representativeness, which is hand-crafted metric compared to the learnable discriminator in HAL. \cite{LiG13} proposed to combine uncertainty with density but only queried one sample a time. The matrix inverse operation in density estimation is expensive for large 3D images. The others integrated generative models, such as adversarial training \citep{zhu2017generative} or variational approaches \citep{SinhaED19} with other heuristics while HAL does not involve any adversarial learning procedure, which would lead to unstable training and data selection results. 

There are also several works that applied active learning for biomedical images \citep{Smailagic0NWKGM18,YangZCZC17,ZhouSZGGL17,KuoHYMM18} that are built upon conventional active learning approaches for natural images. They are either not considering the tradeoff between the two query principles or not applicable for 3D subtomogram classification tasks with high noise and transformation variations, which will lead to a sub-optimal subtomogram selection and classification performance.

\section{Proposed Approach}
Our Hybrid Active Learning method integrates two principles. For the representativeness principle, we develop an alternative optimization scheme for training the multi-class subtomogram classification model and a discriminator. Specifically, given a small initial set of labeled subtomograms, the classification model is firstly trained in a supervised way. Then, the hidden representations of both the labeled and unlabeled subtomogram samples are extracted to train a binary classification model (i.e. the discriminator). Thus, the probability scores of the unlabeled subtomogram samples are obtained from the predictions of the discriminator. Meanwhile, the uncertainty score (i.e. the entropy of the predictions from the multi-class classification model) is further fused with the discriminator score to produce the final query metric for ranking the unlabeled subtomograms. Afterwards, the top subtomograms are selected and labeled for iterative training until the budget is reached.

\subsection{Representativeness Principle}
In this part, the multi-class subtomogram classification model starts with a sparsely labeled dataset $\mathcal{D}$. Within the dataset, we denote the labeled subtomograms at iteration $t$ as $\mathcal{LD}(t)$ and the unlabeled subtomograms as $\mathcal{UD}(t)$. Then we have $\mathcal{LD}(t)\cup \mathcal{UD}(t) = \mathcal{D}$ and $\mathcal{LD}(t)\cap \mathcal{UD}(t) = \emptyset$. From the domain adaptation point of view, we treat $\mathcal{LD}(t)$ and $\mathcal{UD}(t)$ as two separate domains, namely the source domain $\mathcal{L}$ and target domain $\mathcal{U}$ respectively. $M(\cdot)$ is defined as the feature extractor and $D(\cdot)$ is the introduced discriminator which aims to distinguish these two domains. 

At each iteration $t$, the to enhance the sample representativeness, we first train the main classifier using the softmax cross entropy loss. Then we extract the representations from the intermediate layers on both $\mathcal{LD}(t)$ and $\mathcal{UD}(t)$ and regard them as inputs to the discriminator. Next, we train this discriminator by a binary classification task so that it can discriminate the labeled and unlabeled subtomograms well. If we assume the output of the discriminator $D(\cdot)$ to be 0 for labeled class and 1 for unlabeled class, then we select and label a batch of subtomogram samples $B(t)$ which satisfy:
\begin{equation}
\label{eq:represe}
    B(t) = \arg\max_{x \in \mathcal{UD}(t)} \operatorname{Pr}( D(M(x))=1 | M(x)),
\end{equation}
where $B(t)$ is the queried unlabeled batch at iteration $t$.

\textbf{Why a discriminator?} The reason behind is if we can determine with high probability that an unlabeled subtomogram is from $\mathcal{UD}$, then it should be different from $\mathcal{LD}$, which is helpful for improving the information encoded in the labeled dataset and thus better for the model to generalize on the remaining unlabeled subtomogram examples after we label it. Otherwise, if the subtomogram examples from $\mathcal{UD}$ are indistinguishable from $\mathcal{LD}$, then we successfully represent the distribution with $\mathcal{LD}$. This motivates us to design a discriminator $D$ for such probability estimation and alignment. Moreover, the introduced discriminator is expected to provide more flexibility during classification and the subtomogram sample selection since it has a learnable metric for separating the labeled and unlabeled subtomogram examples, which is better than hand-crafted metric designs \citep{ShuiZGW20,TangH19}.

\subsubsection{Task-specific Designs} 

In this section, we proposed two task-specific designs to further refine the capacity of the representativeness principle, namely the convolutional discriminator and the subset sampling strategy. 

\begin{figure*}[!htbp]
	\centering  
	\includegraphics[width=0.85\linewidth]{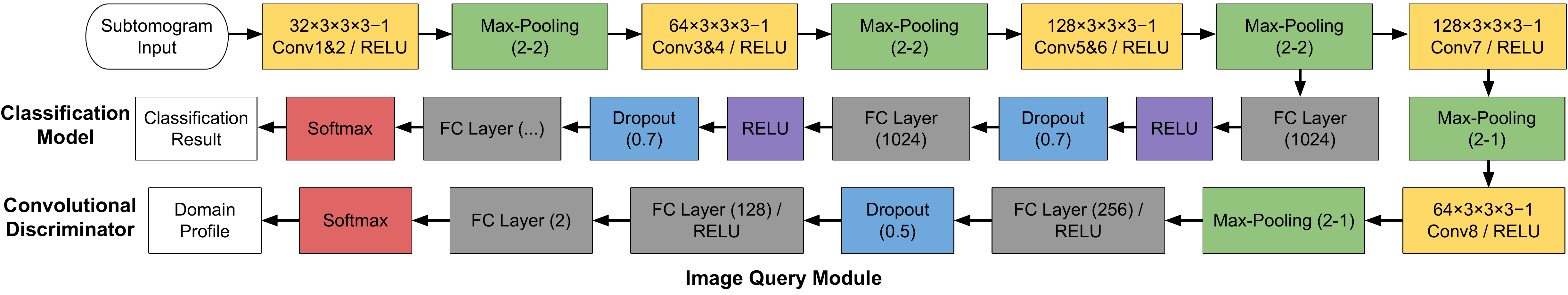} 
	\caption{The model architecture of our subtomogram classifier with detailed layer configuration. ``$64\times3\times3\times3$-1-same" means 64 convolutional filters with $3\times3\times3$ kernel, 1 stride and ``same" padding. ``FC layer (1024)" means 1024 units in the fully connected layer. The shape of the final FC layer depends on the number of classes denoted as ``FC layer (...)". Max-Pooling (2-1) means $2\times2\times2$ filters with 1 stride. Dropout (0.7) means the dropout rate is 0.7 in that layer. ReLU and Softmax denote the activation function.}
	\label{fig:classifier_structure}   
\end{figure*}

Commonly, the discriminator for unsupervised domain adaptation \citep{DBLP:conf/icml/GaninL15} often regards the outputs from the fully connected layers as the input. They claim such design will focus on more fine-grained information for feature adaptation since these layers of the network extract and propagate more specific features. However, these fine-grained features are more suitable for multi-class classification which is usually biased for the discriminator, especially for highly heterogeneous 3D cryo-ET data. Instead, we propose to use the output of the last max-pooling layer as the input for the discriminator and enhance the discriminator with the more flexible convolutional operations. Specifically, the convolutional discriminator consists of two convolutional layers followed by two fully connected layers (Fig.\ref{fig:classifier_structure}). These convolution operations enable our model to learn a flexible representation space and a task-specific comparison metric for binary classification, which is helpful for querying valuable subtomograms more effectively. 

In addition, recall that querying unlabeled subtomogram samples requires iteratively training the multi-class classification model and expert annotation in a loop. Therefore, the querying efficiency is important. One simple solution is to query subtomograms in larger batches instead of one subtomogram at a time, which reduces the waiting time until the classifier finishes training \citep{AzimiFFBH12}. However, since we select the data in a large batch which are all predicted by the discriminator with a high probability to be unlabeled, they tend to have a similar distribution, especially for the subtomograms captured with a higher noise level. This scenario causes significant information overlap. To mitigate this, we emphasize on the diversity of the sampled subtomograms in a batch by assuming consecutive mini-queries will be less likely to contain similar instances. We split the original queried batch $B(t)$ into $m$ sub-batches. Suppose we desire to select $K$ subtomograms at iteration $t$, we first train the discriminator on the representations until convergence and label the top $\frac{K}{m}$ subtomograms. Then we repeat the process by interleaving the discriminator training and subtomogram selection until $K$ subtomogram samples are queried. During this process, we only train the main classifier once but train the discriminator for $m$ times which is more efficient. The detailed architecture of our model is demonstrated in Fig. \ref{fig:classifier_structure}. 
\subsubsection{Theoretical Analysis}
The motivation of the representativeness principle is to label the most appropriate data from the unlabeled subset $\mathcal{UD}$ that can represent the distribution of the training (or the entire) dataset as well as possible. In this case, a classifier trained on $\mathcal{LD}$ should perform similarly compared to that trained with the entire dataset $\mathcal{D}$ labeled. Naturally, we are interested in how to measure the distribution difference between two observations $x \sim \mathcal{L}$ and $x \sim \mathcal{U}$ and see if the design of a discriminator can achieve less classification error on $\mathcal{U}$. Without loss of generality, we use $\mathcal{H} \Delta \mathcal{H}$ divergence \citep{DBLP:conf/vldb/KiferBG04} $d_{\mathcal{H} \Delta \mathcal{H}}\left(\mathcal{L},\mathcal{U}\right)$ for distribution difference estimation, which measures the maximum difference of the probabilities for inconsistent prediction.

Denote $\epsilon_{d}$, $\epsilon_{U}$ and $\epsilon_{L}$ to be the classification error of the discriminator and the multi-class classification model on the unlabeled and labeled subtomogram samples, respectively, we argue $\epsilon_{U}$ is bounded by a term related to $\epsilon_{L}$ and $\epsilon_{d}$  by Theorem 1. 

\textbf{Theorem 1.} \textit{Assume the complexity of the discriminator is more than a XOR function, given $f(\cdot)$ a multi-class candidate classifier, the classification error on the unlabeled dataset is bounded by: $\epsilon_{U}(f) \leq \epsilon_{L}(f)+  \epsilon_{d}+C$. $C$ is an uncorrelated constant. }

\noindent\textit{Proof Sketch.} Following the proof and assumptions made in \citep{DBLP:journals/ml/Ben-DavidBCKPV10} and substituting the source and target domain as $\mathcal{L}$ and $\mathcal{U}$, we get 
\begin{equation}
\label{eq:proof}
    \epsilon_{U}(f) \leq \epsilon_{L}(f)+ \frac{1}{2}d_{\mathcal{H} \Delta \mathcal{H}}\left(\mathcal{L},\mathcal{U}\right)+C.
\end{equation} 
Then following the derivation in \citep{DBLP:conf/icml/GaninL15}, we replace $d_{\mathcal{H} \Delta \mathcal{H}}\left(\mathcal{L},\mathcal{U}\right)$ by its upper bound 
\begin{equation}
    2 \sup _{\eta \in \mathcal{H}_{d}}\left|\operatorname{Pr}_{\mathcal{L}}[z: \eta(z)=1]+\operatorname{Pr}_{\mathcal{U}}[z: \eta(z)=0]-1\right|,
\end{equation}
which can be seen as $2\epsilon_{d}$. Here $\mathcal{H}_{d}$ is the function space for the discriminator. And after substitution,
the theorem is proven. Here the assumption is easily satisfied since the discriminator is implemented by a neural network which is complex enough according the Universal Approximation Theorem \citep{Barron93,Funahashi89}.

Given such a guarantee, if $\epsilon_{L}(f)$ and $\epsilon_{d}$ is minimized, the classification error on the unlabeled dataset is bounded.

\subsection{Discriminativeness Principle}
In addition to the introduced discriminator for improving the sample representativeness, we argue that the useful label information (i.e. inductive bias) is missing in the current query strategy, which is shown to be effective in literature \citep{WangZLZL17,Gal2017Deep,BeluchGNK18}. Thus, we propose a hybrid query method by selecting discriminative subtomogram samples with uncertainty sampling \citep{Yang2018A}.

The intuition is as follows: the representativeness principle assumes the unlabeled pool is large enough to represent the true distribution. However, the data from the sparse regions of distribution will be sampled because the unlabeled set gradually becomes not representative due to its decreasing size. Conversely, uncertainty sampling can keep a balance between labeled and unlabeled subtomograms on the representation space by selecting subtomograms corresponding to data density such that the classification will not easily be biased by the sparse region of the manifold \citep{Yang2018A}. 

On the other hand, uncertainty sampling is designed to sample the most uncertain instance which is closest to the decision boundary. Since the number of subtomograms in the initial stage is limited, the estimated decision boundary is far from the actual one. Therefore, it may select noisy instances and stuck at sub-optimal solutions due to a lack of exploration. In contrast, the semi-supervised setting in the discriminator-based query strategy can avoid this drawback by observing the entire dataset. Therefore, during training, the representativeness principle by the discriminator-based query and the discriminativeness principle by uncertainty sampling assist each other and further enhance the stability of query and classification performance. 

Specifically, we use the entropy of the predictions to measure the uncertainty, which is formulated as:
\begin{equation}
\label{eq:uncertainty}
    E(t)=\arg \max _{x \in \mathcal{UD}(t)} \left[-\sum_{y \in C} P(y | x) \log P(y | x) \right],
\end{equation} 
where $C$ denotes the class space. $P(y | x)$ denotes the conditional probability of $y$ given $x$ in the multi-class classifier. We evaluate the quality of the model prediction $P(y|x)$ by comparing the class with the highest probability against the ground truth labels given by domain experts. We implement this evaluation in the neural networks by using the softmax loss function, which is the common practice in image classification. 

\subsection{Hybrid Active Learning}
In order to tradeoff between the two principles, at each iteration, we design a ranking score for final selection criteria:
\begin{equation}
\label{eq:final_criteria}
    S(t)=\operatorname{Pr}( D(M(x))=1 | M(x))+\lambda E(t),
\end{equation}
where $\lambda$ is the weighting hyperparamter for balancing different scores. The other notations have the same meaning as Eqn. \ref{eq:represe} and \ref{eq:uncertainty}. While there remain other score fusion methods, we argue our implementation is simple and effective enough to achieve sufficient application purposes.

\section{Experiments and Results}
\subsection{Dataset and Preparation}
We evaluate our method on two simulated and three real cryo-ET datasets. For simulation datasets, we utilize the PDB2VOL program \citep{WRIGGERS1999185} to generate 23 classes of subtomograms which have the same class space as \citep{DBLP:journals/bioinformatics/XuCMLYZX17} at two Signal-to-Noise Ratio (SNR) levels, including $0.03$ (S1) and $0.05$ (S2). These datasets are realistically simulated by
approximating the true cryo-ET image reconstruction process through a tilt-angle of $\pm \ang{60}$, 
including the Contrast Transfer Function and Modulation
Transfer Function. 
Each class contains $1,000$ subtomograms with size of $40^{3}$ voxels. These simulated datasets are used in our 23-class classification tasks. 

For real datasets, we use a set of rat neuron tomograms from \citep{Guo2018In} (R1). For one tomogram, we manually select 1,800 subtomogram samples which contain particles of $28^{3}$ voxels from 5424 subtomograms extracted by Difference of Gaussian (DoG) \citep{Long2016Simulating} (R1a). We evaluate the particle picking task for determining whether or not a sample contains a particle. This is formulated as a binary classification task for the multi-class classification model. We also extract 2,394 subtomograms with size of $40^{3}$ in the same tomogram set. The 2,394 subtomograms contain 6 classes detected and classified by template matching (R1b) \citep{Guo2018In}. We evaluate the 6-class classification task on it. In addition, we process a 7-class dataset \citep{Noble230276} (R2) from EMPIAR \citep{iudin2016empiar}. Each class contains $400$ subtomograms with size of $28^{3}$. Following common practice in the active learning literature, no data augmentation techniques are used during training. The effect of data augmentation remains to be further explored.

\begin{figure}[!ht]
	\centering  
	\includegraphics[width=0.8\linewidth]{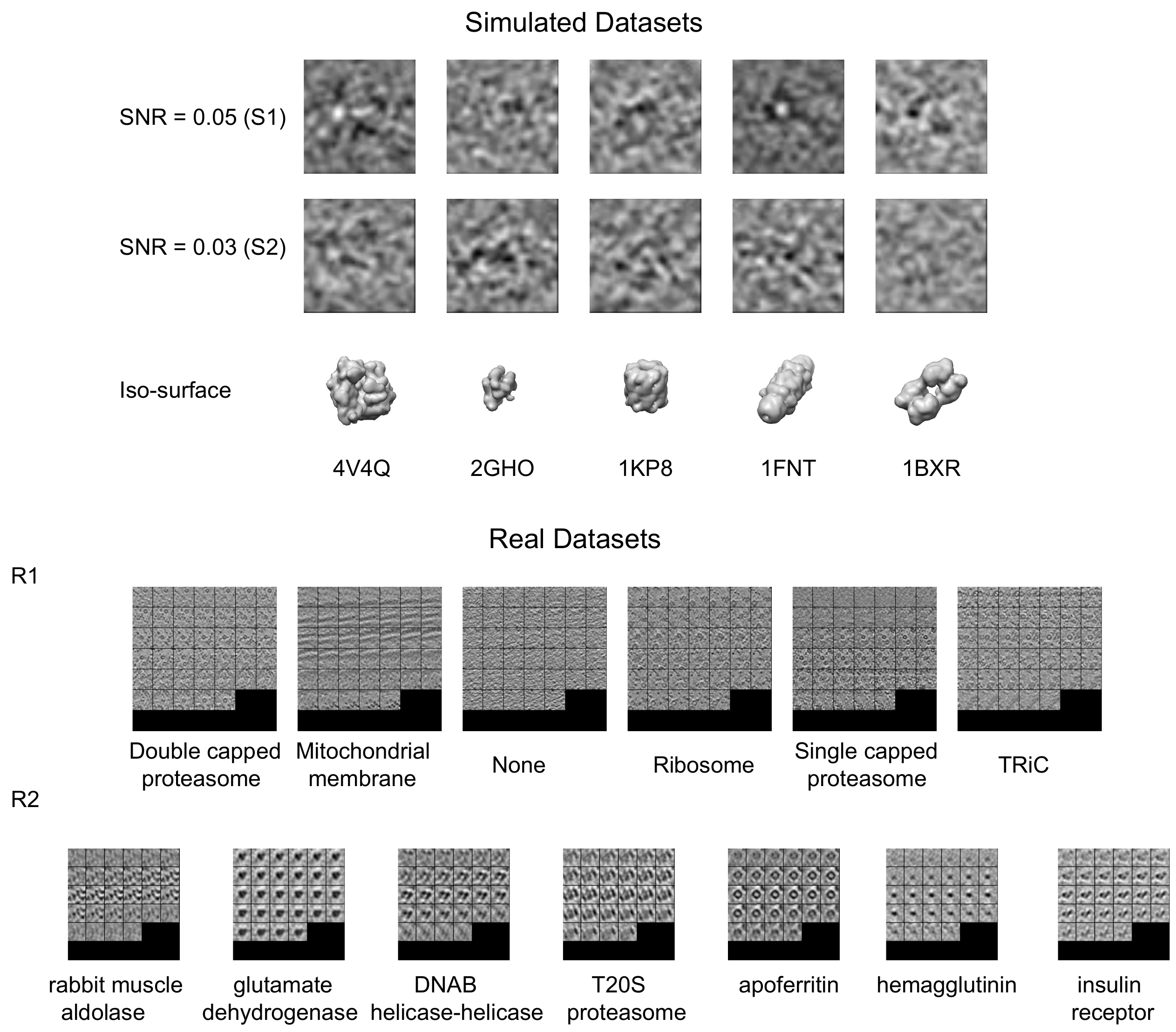}  
	\caption{Examples of used subtomograms. For simulated datasets, 5 out of 23 classes of simulated subtomograms are shown for simplicity. We plot in the form of iso-surface (bottom row) and center-sliced density map in parallel with the x-z plane (the first two rows). Their PDB IDs are below each image. For real datasets, we visualize the center-sliced density map in parallel with the x-z plane with the name of each macromolecule class below each image.
 }	
	\label{fig:dataset}   
\end{figure}

A 7-class classification task is evaluated on R2. The 2D $x-z$ center slice of the 3D images and the iso-surface of the simulated datasets are demonstrated in Fig. \ref{fig:dataset}.

\subsection{Results and Comparisons}
In this section, we report the subtomogram classification result for both the simulated and real data. We start with $3\%$ of the entire dataset as the labeled subtomogram samples for the simulated datasets $S1,S2$ and $3\%, 4\%, 4\%$ for real datasets $R1a,R1b$ and $R2$, respectively. The effect of the number of the initially labeled subtomogram samples is shown in the next section. The query batch size is empirically fixed to 800 for the simulated datasets and 32 for the real datasets, which follows the common active learning setting \citep{tran2019bayesian}. In terms of the subset sampling for the simulated datasets, we report the model performance with the number of subset to be 20 while the number is $4, 8, 4$ for the real datasets $R1a, R1b$ and $R2$, respectively. The effect of the number of the subset and subset size is deferred for discussion in the next section. We report the classification results after 7,5,6,5 and 8 query iterations in dataset S1, S2, R1a, R1b and R2, respectively since we empirically found more iterations will not bring significant improvement on HAL. We run all the baselines under the same setting and report all metrics using an average of 10 runs with random seed from 1 to 10.

In Tab.\ref{tab:comp_baselines}, we firstly compare with supervised training with the entire dataset labeled. In dataset S2, we use $16.91\%$ labeled training data to achieve $93.86\%$ test accuracy compared to $95.36\%$ in fully supervised training. In dataset R1a, we use $11.89\%$ of training data to achieve $85.48\%$ test accuracy compared to $87.24\%$ in fully supervised training.  Moreover, we compared with 8 representative active learning baselines, including methods using a single query principle, namely, Random Query \citep{DBLP:conf/iscide/WooP12}, Uncertainty Query \citep{JoshiPP09}, CoreSet Query \citep{Sener2018Active}, Bayesian Query \citep{Gal2017Deep}, Bayesian Generative Active Learning \citep{tran2019bayesian} (BGAL) and hybrid query heuristics, exploration-exploitation BMAL \citep{YinQCLWZD17} (EE-BMAL), VAAL \citep{SinhaED19} and BADGE \citep{AshZK0A20}. As shown in Tab.\ref{tab:comp_baselines}, our method achieves a superior performance on all 5 different datasets under the same labeling budget. For single query principle, BGAL performs the best compared to others, especially on dataset R1a, which achieves a $82.18\%$ final accuracy. Surprisingly, even with a theoretical guarantee, the core-set sampling performs the worst among baselines, which is possibly caused by the complex data distribution that makes it harder to cover the entire dataset with the constructed core-sets. Moreover, the baselines that adopt a hybrid query strategy usually performs better because of the mutual benefits of different criteria. However, they still underperforms our HAL which explicitly trade-offs the representativeness and the discriminativeness principle. 
\begin{table*}[!t]
\centering
\small
\begin{tabular}{c|ccccc}
\hline
  Method/Dataset & S1 & S2 & R1a & R1b & R2\\ 
  \hline
Supervised Training & $83.68^{\pm0.24}$&$95.36^{\pm0.11}$&$87.24^{\pm0.09}$&$81.53^{\pm0.37}$&$99.00^{\pm0.02}$\\
   HAL & $ \mathbf{80.23^{\pm1.47}}$ &  $\mathbf{93.96^{\pm0.42}}$&  $\mathbf{85.48^{\pm0.56}}$  &   $\mathbf{74.80^{\pm0.33}}$ &  $\mathbf{95.00^{\pm0.94}}$  \\
 Random Query \citep{DBLP:conf/iscide/WooP12}  & $ 74.77^{\pm2.34}$ &  $77.66^{\pm0.98}$  &  $67.00^{\pm1.13}$  &  $67.30^{\pm0.22}$ & $ 85.85^{\pm1.53}$ \\
 Uncertainty Query \citep{JoshiPP09} & $77.32 ^{\pm1.00}$  & $ 90.78^{\pm1.21}$  &  $66.29^{\pm0.80}$& $ 70.00^{\pm0.75}$ &   $73.00^{\pm0.28}$ \\
 Bayesian Query \citep{Gal2017Deep} &$ 73.23 ^{\pm2.64}$ & $ 85.29^{\pm1.99}$  &   $77.00^{\pm0.89}$&  $70.70^{\pm0.74}$  &   $78.00^{\pm0.55}$\\
 CoreSet Query \citep{Sener2018Active} & $63.59 ^{\pm2.62}$  &  $63.48^{\pm1.88}$  &   $78.17^{\pm0.42}$ &  $62.60^{\pm0.59}$ &   $42.00^{\pm0.92}$ \\
 BGAL \citep{tran2019bayesian} &$78.23^{\pm0.65}$& $85.32^{\pm1.01}$&$82.18^{\pm0.99}$& $71.22^{\pm1.33}$&$88.34^{\pm0.93}$ \\
 VAAL \citep{SinhaED19}& $75.67^{\pm0.91}$& $86.51^{\pm1.23}$& $83.33^{\pm1.08}$& $69.72^{\pm0.86}$&$85.22^{\pm0.85}$\\
 EE-BMAL \citep{YinQCLWZD17} &$79.32^{\pm0.77}$&$89.11^{\pm1.35}$&$84.34^{\pm1.00}$&$71.62^{\pm2.01}$&$90.67^{\pm1.05}$ \\
 BADGE \citep{AshZK0A20} &$79.46^{\pm0.41}$&$91.00^{\pm0.73}$&$82.63^{\pm0.81}$&$73.01^{\pm0.54}$&$93.21^{\pm1.11}$ \\
 \hline
 Labeled Percentage & $23.87\%$&$16.91\%$ &$11.89\%$ &$9.35\%$&$12.00\%$ \\
 \hline
\end{tabular}
\caption{Comparison of HAL and the baseline AL methods on five different datasets (results are the classification accuracy in \%). The same labeling budget is used among different methods. The standard deviation is reported at the top right corner.}
\label{tab:comp_baselines}
\end{table*}

\begin{table*}[!t]
    \centering
    \small
    \tabcolsep 0.04in\renewcommand\arraystretch{0.545}{\small{}}
    \begin{tabular}{ccc|ccc|ccc|ccc|ccc}
    \hline
    \multicolumn{3}{c}{S1} & \multicolumn{3}{c}{S2} & \multicolumn{3}{c}{R1a} & \multicolumn{3}{c}{R1b} & \multicolumn{3}{c}{R2}\\
    \hline
    Config& Acc & T  &Config &  Acc & T & Config & Acc & T & Config & Acc & T & Config &Acc & T \\
    \hline
    400/2 &  75.96 & 1.9 h &  400/2 & 83.62 &  1.6 h & 32/1&   80.25 &  0.2 h &32/1&    63.50 &  0.2 h&32/1 & 90.86 &  0.4 h \\
    80/10 & 77.96& 3.8 h& 80/10& 85.20& 2.7 h& 16/2 &81.23 & 0.4 h&16/2& 71.74& 0.4 h & 16/2& 90.69& 0.4 h\\
    40/20 &    \textbf{80.23 }&  5.4 h &40/20 &   \textbf{93.96} &  4.6 h &8/4 &    \textbf{85.48} &  0.5 h &8/4& 70.08 &  0.4 h &8/4 &    \textbf{95.00} &  0.5 h\\
      20/40 & 78.82  & 8.0 h & 20/40& 86.65& 6.6 h&4/8 & 83.34& 0.6 h& 4/8& \textbf{74.80}& 0.5 h&4/8& 94.11& 0.6 h \\
    10/80 & 78.56 & 9.1 h& 10/80& 89.73& 8.4 h&2/16 & 84.62& 0.9 h&2/16& 74.09&0.7 h&2/16&93.28&0.7 h\\
      2/400  & 79.49 &  14.7 h &2/400 &    91.08 & 10.6 h &1/32 &    80.19 &  1.0 h &1/32 &    73.21 &  0.8 h &1/32 &   88.62 &  0.7 h\\
      \hline
    \end{tabular}
    \caption{Comparative Results of different subset configurations on HAL (in \%). T refers to the overall training time. Config is in the form of Subset Size/Number of Subset. h denotes hours.}
    \label{tab:combination}
\end{table*}
\subsection{Ablation Study}
To validate the effect of our task-specific designs and the hybrid query strategy, we did a controlled experiment that removes the convolutional layers in the discriminator (Variant 1), the subset sampling (Variant 2) and the uncertainty sampling strategy (Variant 3), which is shown in Tab.\ref{tab:abaltion_new}. Note that we did not remove the representativeness principle because that degenerates the model to the baseline method of Uncertainty Query. The training setting is the same as the previous section.
\begin{table}[!t]
    \centering
    \small
    \begin{tabular}{c|ccccc}
    \hline
       Model/Dataset  &  S1 & S2 & R1a & R1b & R2\\
       \hline
       Variant 1 (V1)& $78.45$ &  $91.87$&  $83.65$  &   $72.27$ &  $94.28$\\
       Variant 2 (V2)&$ 75.77$ &  $91.92$   &  $80.25$  &  $63.50$ &  $90.86$\\
       Variant 3 (V3)&$75.32$ &  $88.63$&  $79.99$  &   $70.40$ &  $91.38$\\
       HAL & $\mathbf{80.23}$ &  $\mathbf{93.96}$&  $\mathbf{85.48}$  &   $\mathbf{74.80}$ &  $\mathbf{95.00}$\\
         \hline
    \end{tabular}
    \caption{Ablation study results (In $\%$) on the convolutional discriminator, subset sampling and the uncertainty sampling}
    \label{tab:abaltion_new}
\end{table}
According to Tab.\ref{tab:abaltion_new}, removing any of the three parts will lead to performance drop. For example, removing the subset sampling will destabilize the training which decreases the accuracy from $74.80\%$ to $63.50\%$ on $R1b$.
\subsection{The effect of initially labeled subtomograms}
To observe the effect of the number of initially labeled subtomogram samples, we test our HAL under different ratios from $1\%$ to $5\%$ with the interval of $1\%$ on five datasets while keeping the other training configuration unchanged. The comparative result is shown in Tab.\ref{tab:initial_ratio}.
\begin{table}[!t]
    \centering
    \small
    \begin{tabular}{c|ccccc}
    \hline
       Ratio/Dataset  &  S1 & S2 & R1a & R1b & R2\\
       \hline
       0.01  & $77.24$ &  $89.99$&  $72.13$  &   $43.61$ &  $88.32$\\
       0.02& $ 80.01$ &  $93.36$   &  $81.62$  &  $47.77$ &  $93.29$\\
       0.03&$\mathbf{80.23}$ &  $\mathbf{93.96}$& $\mathbf{85.48}$  &$73.95$ & $94.37$\\
       0.04 &  $80.96$ & $94.01$ &$84.93$ &  $\mathbf{74.80}$ &  $\mathbf{95.00}$ \\
       0.05& $81.11$& $93.99$&$86.05$&$74.95$&$95.23$\\
         \hline
    \end{tabular}
    \caption{Comparative results (In $\%$) on the number of initially labeled subtomograms for five datasets.}
    \label{tab:initial_ratio}
\end{table}


As shown in Tab.\ref{tab:initial_ratio}, the number of the initially labeled subtomogram samples have considerable effects on the final classification accuracy. The model is trained towards a sub-optimal direction if this ratio is too small, leading to much lower accuracy even though more subtomogram samples are labeled in the later stages. However, the initially labeled subtomogram samples are practically expensive to obtain. Thus, observing that the final accuracy does not increase too much if we keep increasing the number of initially labeled subtomogram samples, we empirically fix the ratio to be $3\%$ for dataset $R1a,S1,S2$, and $4\%$ for dataset $R1b,R2$.
\subsection{The effect of subset configuration}
During subset sampling, it is important to determine the optimal combination of the number of subsets $m$ and the subset size $\frac{K}{m}$ in order to balance the training time and the diversity of the subset. Specifically, we test 6 different subset size for five datasets whose results are summarized in Tab.\ref{tab:combination}. As can be observed, a balanced configuration of subset size and numbers can help achieve better performance. Meanwhile, the time complexity increases dramatically if the subset size is smaller since much more iterations are trained on the discriminator. Therefore, to balance time and accuracy, a moderate subset size is preferable.
\subsection{Query Visualization}
\begin{figure*}[htbp]
	\centering 
 	\includegraphics[width=0.8\linewidth]{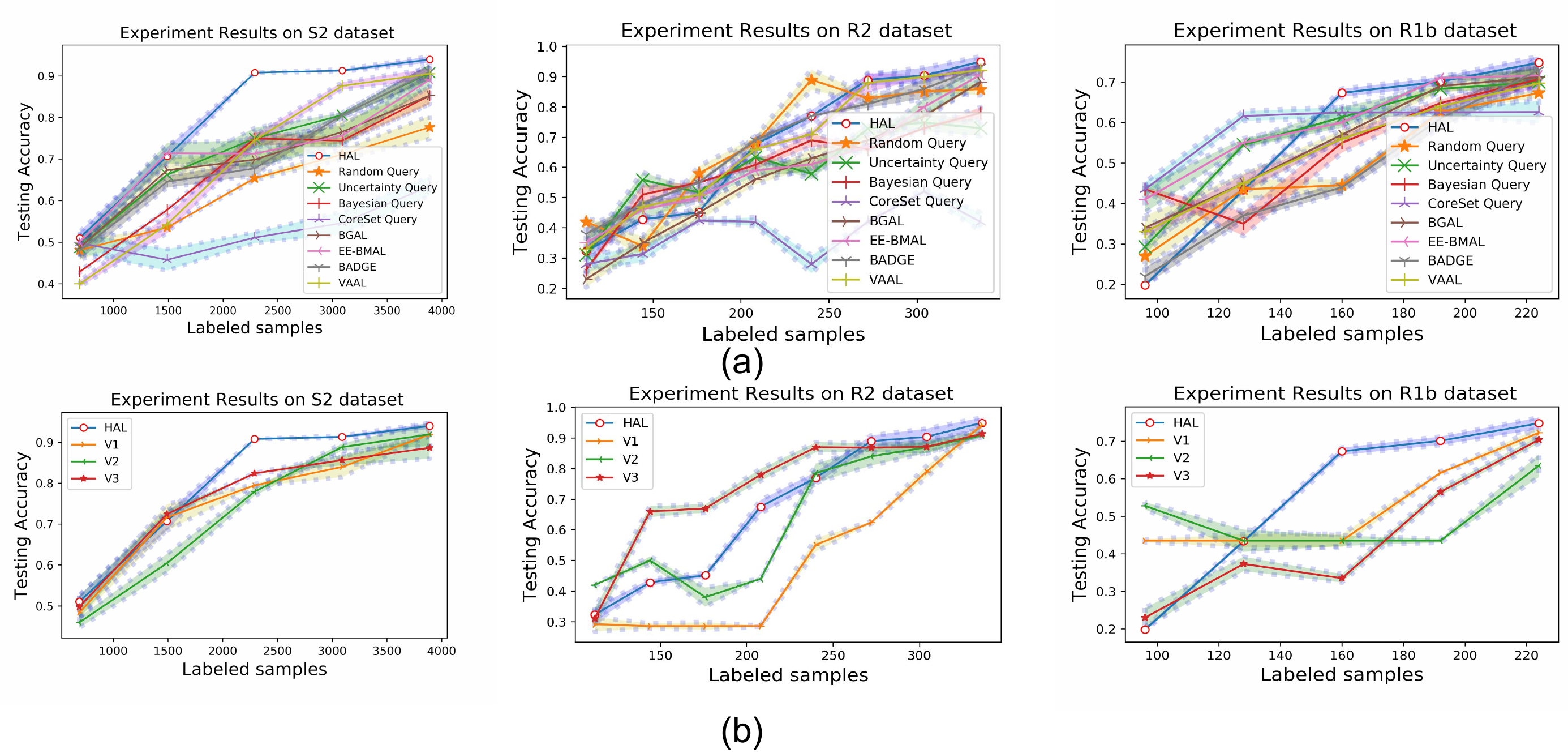}  
	\caption{Comparative querying process with baselines (a) and ablations (b). The shaded area means the standard deviation. For simplicity, three of five datasets are shown.}  
		\label{fig:results_baselines}  
 \end{figure*}

For a comprehensive discussion, we plot the comparative query process in Fig.\ref{fig:results_baselines}. Here we demonstrate the accuracy versus the number of labeled subtomogram samples during training. We can see the stability and the final classification accuracy of HAL is better without accuracy decrease or stagnation along the sampling procedure.


\subsection{The effect of $\lambda$}
\label{sec:effect_lambda}
HAL relies on the representative score predicted by a discriminator and the discriminative score calculated from the entropy of classification to sample examples for labeling. To determine a relatively optimal balance between the entropy and the representative scores, we test five different values of $\lambda$ in Eqn.~5 and report the classification accuracy under the same labeling budget. From Tab.~\ref{tab:ablation_lambda}, we find that $\lambda=1$ obtains a consistently better performance on the dataset S1 and R1a. Increasing or decreasing $\lambda$ degenerates the HAL to the model solely trained with one score, either the representative score or the discriminative score. This set of experiments illustrates the importance of selecting a proper $\lambda$ in order to guarantee a good test performance. 

\begin{table}[!ht]
    \centering
    \begin{tabular}{c|cc}
    \hline
      $\lambda$ & Acc. on S1 & Acc. on R1b\\
        \hline
     0.1  & 79.00&83.26 \\
         0.5& 81.11&84.23  \\
          1& \textbf{80.23}&\textbf{85.48} \\
             2  & 80.94&85.32  \\
             3   & 79.26&84.73  \\
         \hline
    \end{tabular}
    \caption{Ablation study results on $\lambda$. Classification accuracy on dataset S1 and R1b is reported.}
    \label{tab:ablation_lambda}
\end{table}

\subsection{Implementation Details}
Firstly, we normalize the data in every dataset. For classification on both simulated and real data, we randomly split the entire dataset to the train and test set by a ratio of 3:1. We set the learning rate and the batch size of the multi-class subtomogram classifier and the discriminator as 0.001, 128 and 0.01, 256 without further fine-grained tuning. We train the discriminator with early stopping when the accuracy reaches 98\% in order to prevent overfitting. 
we set the $\lambda$ in Eqn.~5 to 1 because both scores have the same value range. Code is available at\footnote{https://github.com/xulabs/aitom}.

\section{Conclusion}
Computational analysis, deep learning approaches in particular, has played an increasingly important role for obtaining molecular machinery insights from cryo-ET data. However, the heavy labeling work behind data-driven methods presents obstacles for biologists to use them as assistant approaches. In this paper, we present a novel active learning tool in the cryo-ET domain with concerns for limited labeling resources, which approaches the active learning objective by querying both representative and discriminative subtomogram samples. Our experimental results on both simulated and real data demonstrate it produces significantly improved test performance compared to baselines under the same labeling budget. Our method represents an important step towards fully utilizing deep learning for \emph{in situ} recognition of macromolecules inside single cells captured by cryo-ET. It can potentially also be very useful for other biomedical research with limited labeling resources.

\section*{Funding}

This work was supported in part by U.S. National Institutes of Health (NIH) grants P41GM103712, R01GM134020, and K01MH123896, U.S. National Science Foundation (NSF) grants DBI-1949629 and IIS-2007595, Mark Foundation For Cancer Research 19-044-ASP, and AMD COVID-19 HPC Fund. XZ was supported by a fellowship from Carnegie Mellon University's Center for Machine Learning and Health.

\section*{Author information}
Statement on authorship changes: Dr. Eric Xing was an academic advisor of Mr. Haohan Wang. Dr. Xing was not directly involved in this work and has no direct interaction or collaboration with any other authors on this work. Therefore, Dr. Xing is removed from the author list according to his request. Mr. Zhenxi Zhu participated in this work in early 2019 as a small part of his undergraduate research training. His affiliation is updated to his current affiliation.


\bibliographystyle{natbib}
\bibliography{arxiv}

\end{document}